\documentclass{aip-cp}

\usepackage[numbers]{natbib}
\usepackage{rotating}
\usepackage{graphicx}
 \usepackage{epstopdf}
\begin{document}

\title{Measurements of Hadron Form Factors at BESIII}

\author[aff1]{Cristina Morales Morales\corref{cor1} on Behalf of BESIII Collaboration}


\affil[aff1]{Helmholtz-Institut Mainz, 55099 Mainz, Germany.}
\corresp[cor1]{Corresponding author: c.morales-morales@gsi.de}

\maketitle

\begin{abstract}
BEPCII is a symmetric $e^+e^-$-collider located in Beijing  running at center-of-mass energies between 2.0 and 4.6 GeV. This energy range allows the BESIII-experiment to measure hadron form factors both from direct $e^+e^-$-annihilation and from initial state radiation processes. In this paper, results on $e^+e^-\rightarrow p\bar{p}$ based on data collected by BESIII in 2011 and 2012 are presented. We also present preliminary results on $e^+e^-\rightarrow \Lambda \bar{\Lambda}$ based on the same data samples at 4 center-of-mass energies. BESIII results obtained from $e^+e^-\rightarrow \pi^+\pi^-$ using the initial state radiation technique at the center-of-mass energy of 3.773 GeV are also summarized.
Finally, expectations on the measurement of baryon electromagnetic form factors from the BESIII high luminosity energy scan in 2015 and from initial state radiation processes at different center-of-mass energies are also reported. 

\end{abstract}

\section{INTRODUCTION}
Form factors (FFs) account for the non point-like structure of hadrons. Depending on its spin, s, a hadron has 2s+1 form factors. The FFs are analytic functions of the momentum transferred by the virtual photon, $q$. They are real in the space-like region ($q^2<0$) and complex in the time-like region ($q^2>0$) for $q^2 > 4m_\pi^2$. FFs at $q^2<0$ are determined by elastic scattering of electrons from hadrons available as targets. FFs at  $q^2>0$ are measured in annihilation  processes \mbox{$e^+e^- \leftrightarrow h \overline{h}$}.

Spin 1/2 baryons have two electromagnetic FFs. Here the Sachs FFs in the form of $G_E$ and $G_M$ are used.
The Born differential cross section of the $e^+e^-$-annihilation into a baryon-antibaryon pair, \mbox{$e^+e^- \rightarrow B\overline{B}$}, in the $e^+e^-$ center-of-mass (c.m.) reads~\cite{Zichichi} 
\begin{equation}
\label{diff}
\frac{d\sigma^{\mathrm{Born}}(q^2,\theta_{B}^*)}{d\Omega} = \frac{\alpha^2\beta C}{4q^2} \left [(1 + \mathrm{cos}^2\theta_{B}^*) |G_M (q^2)|^2 +  \frac{1}{\tau}\mathrm{sin}^2\theta_{B}^*|G_E(q^2)|^2 \right ],
\end{equation}
where $\theta_{B}^*$ is the polar angle of the baryon, $\tau = 4m^2/q^2$ , $m$ the baryon mass, and $\beta = \sqrt{1-1/\tau}$. The Coulomb factor,  $C = y/(1-\mathrm{exp}(-y))$ with $y = \pi \alpha / \beta$, accounts for the electromagnetic $B \overline{B}$ interactions of point-like baryons~\cite{Tzara} and is equal to 1 for neutral baryon pairs. Angular integration of the previous equation gives the total cross section:
\begin{equation}
\sigma^{\mathrm{Born}}(q^2) = \frac{4\pi\alpha^2\beta C}{3q^2} \left[ |G_M(q^2)|^2 + \frac{1}{2\tau} |G_E(q^2)|^2 \right ]. 
\label{totalcs}
\end{equation}
Experiments usually quote measurements of $|G_M(q^2)|$ under the working hypothesis $|G_E(q^2)| = |G_M(q^2)|$. By defining the effective form factor (EFF) as 
\begin{equation}
|G(q^2)|^2 =  \frac{2\tau |G_M(q^2)|^2 + |G_E(q^2)|^2}{2\tau +1} = \frac{\sigma^{\mathrm{Born}}(q^2) }{(1+\frac{1}{2\tau})(\frac{4\pi\alpha^2\beta C}{3q^2})} , 
\label{effectiveff}
\end{equation}
it is possible to compare with older measurements. The simultaneous extraction of $|G_E|$ and $|G_M|$ without assumptions is only possible by measuring the angular distributions of the outgoing particles (Eq.~\ref{diff}).

The process of $e^+e^-$-annihilation can be accompanied by the emission of one or several high energy photons from the initial state (ISR).
The following equation relates the differential cross section of the ISR process
and the cross section of the direct annihilation:

\begin{equation}
\frac{d^2\sigma^{\mathrm{ISR}}} {dq^2d\theta_{\gamma}^*} = \frac{1}{s} \cdot W(s,x,\theta_{\gamma}^*) \cdot \sigma^{\mathrm{Born}}(q^2),
\label{differential}
\end{equation}
where $x = 2E_\gamma^*/\sqrt{s} = 1 - q^2/s$, $\sqrt{s}$ is the c.m. energy of the collider, and $E_\gamma^*$ and $\theta_\gamma^*$ are the energy and polar angle of the ISR photon in c.m. The radiator function $W(s,x,\theta_{\gamma}^*)$ describes the probability of the ISR photon emission:
\begin{equation}
W(s,x,\theta_{\gamma}^*) = \frac{\alpha}{\pi x} \cdot  \left(\frac{2-2x+x^2}{\mathrm{sin}^2\theta_\gamma^*} - \frac{x^2}{2}\right ), 
\label{radiator}
\end{equation}
with $\alpha$ the electromagnetic fine structure constant~\cite{Radiator}. The emission of the ISR-photon at very small or very large polar angle in c.m. and with low energies is favored. Depending on the energy of the ISR photon, the hadronic invariant mass in the final state is reduced and the hadronic cross section can be extracted for all masses below the actual c.m. energy of the collider up to the production threshold of the hadronic state.

\section{THE BESIII EXPERIMENT AND DATA SETS}
BEPCII is a double ring $e^+e^-$ symmetric collider running at $\sqrt{s}$ from 2.0 to 4.6 GeV. The design luminosity is $1 \times 10^{33}$~$\mathrm{cm}^{-2}\mathrm{s}^{-1}$ at a beam energy of 1.89 GeV and has been achieved in 2016. BESIII is a cylindrical detector which covers 93\% of the full solid angle~\cite{bes3}. It consists of the following sub-detectors: a Multilayer Drift Chamber (MDC);
a Time-of-Flight plastic scintillator (TOF); 
a CsI(Tl) Electro-Magnetic Calorimeter (EMC);
a superconducting magnet of 1T
and finally a Muon Chamber (MUC).
BESIII has accumulated the world$^\prime$s largest samples of $e^+e^−$-collisions in the tau-charm region. Furthermore, in 2015 BESIII peformed a high luminosity scan in 22 energy points between $\sqrt{s}=$ 2.0 and 3.08 GeV, with about 650  $\mathrm{pb^{-1}}$ luminosity. These statistics are the highest in this energy region and, in combination with the previously collected scan data between 2.23 and 4.6 GeV (1.3 $\mathrm{fb^{-1}}$), can be used for the study of $e^+e^- \rightarrow \pi^+\pi^-, N \bar{N}, \Lambda \bar{\Lambda}, \mathrm{etc.}$ In addition, the BESIII data collected at different charmonium and XYZ-states above 3.77 GeV (7.4$~\mathrm{fb}^{-1}$ before 2016) can be used for the study of the corresponding ISR channels.

\section{NUCLEON FORM FACTORS}
\begin{figure}[t]
\begin{minipage}{16.5pc} \hspace{1cm}
\includegraphics[width=16.5pc]{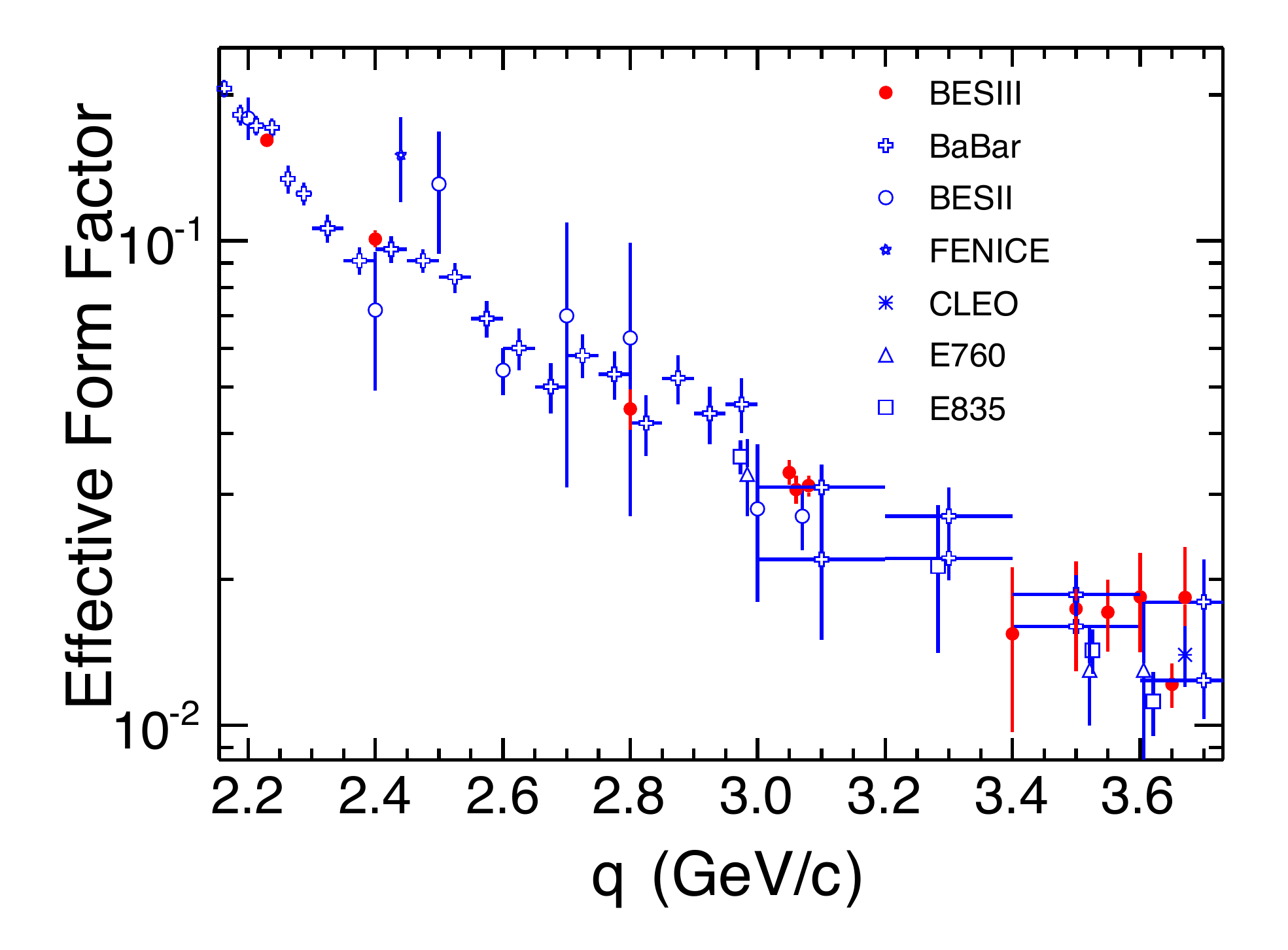}
\caption{dagv<bvy} \label{FFxiaorong}
\end{minipage}\hspace{2pc}%
\begin{minipage}{16.5pc}
\includegraphics[width=16.5pc]{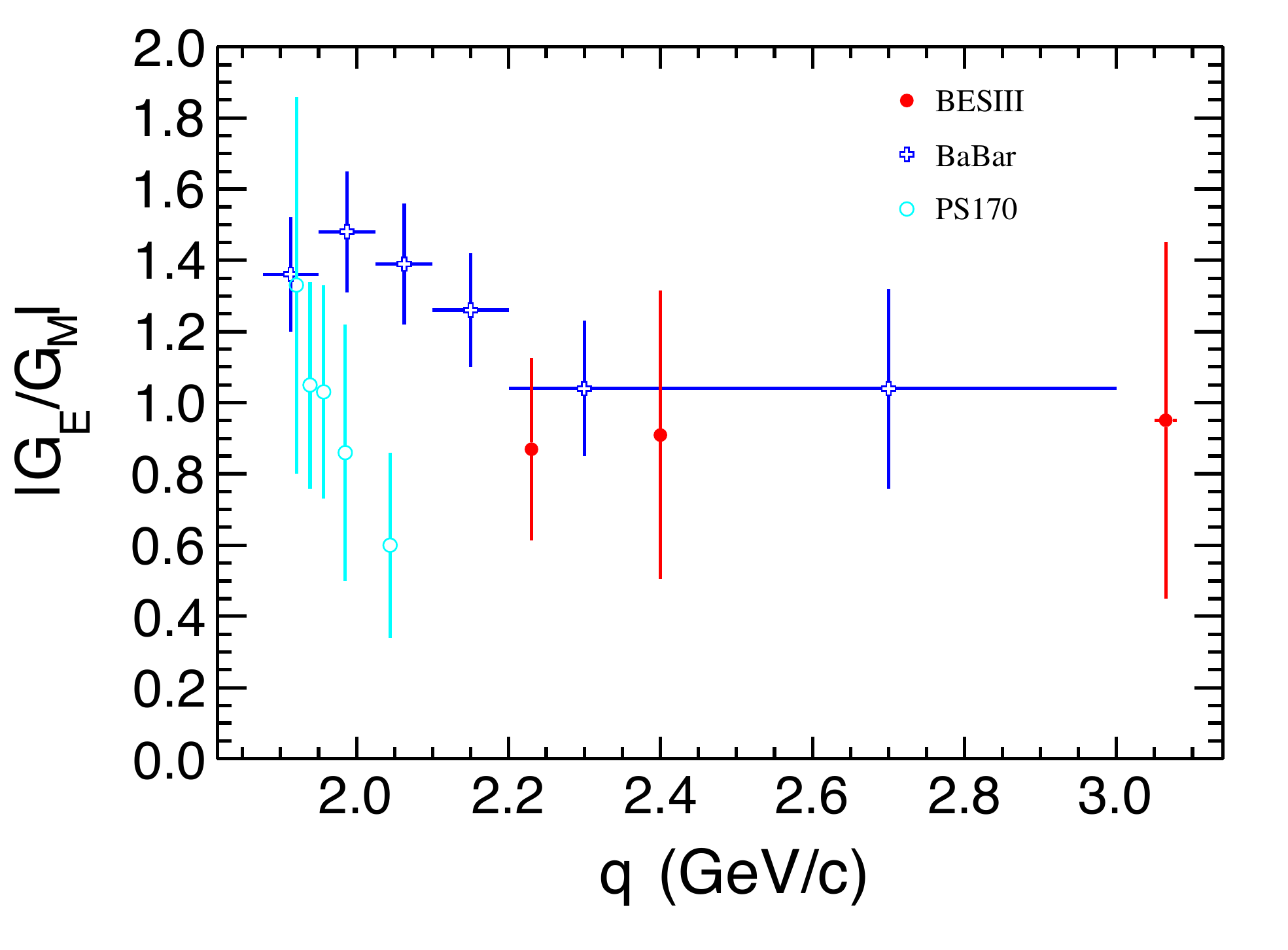}
\caption{Proton effective form factor (left) and ratio of proton electro-magnetic form factors (right) in the time-like region. Reprinted from~\cite{xiaorong}.}
\end{minipage}
\end{figure}

In June 2015, BESIII published the measurement of the channel $e^+e^-\rightarrow p\bar{p}$ at 12 c.m. energies ranging from 2.2324 to 3.6710 GeV~\cite{xiaorong}. These data were collected in 2011 and 2012 and correspond to a luminosity of 157 $\mathrm{pb^{-1}}$. The Born cross section was extracted according to 
\begin{equation}
\sigma_\mathrm{Born} =  \frac{N_{obs} - N_{bkg}}{\mathcal{L}\cdot\epsilon(1+\delta)}, 
\label{xsmeasured}
\end{equation}
where the number of background events is subtracted from the observed signal event candidates, normalized with the luminosity at each scan point, $\mathcal{L}$, and corrected with the selection efficiency of the process, $\epsilon$,  times the ISR radiative correction factor up to next-to-leading order (NLO) $(1+\delta)$. The ConExc generator~\cite{conexc} was used both for the efficiency and the radiative factor evaluation. 
The accuracy in the cross section measurements was between 6.0$\%$ and 18.9$\%$ up to $\sqrt{s} < 3.08$ GeV. The EFF was extracted according to Eq.~\ref{effectiveff} and is shown in Fig.~\ref{FFxiaorong} (left) together with previous experimental results~\cite{bes,CLEO,BABARpp,PS170,E760,E835,cmd3}. A fit to the polar angular distribution of the proton in c.m. was performed according to Eq.~\ref{diff} and the the ratio $|G_E$/$|G_M|$ and $|G_M|$ were extracted for three energy points. The results for $|G_E|$/$|G_M|$ are shown in Fig.~\ref{FFxiaorong} (right). An expansion in moments of the angular distribution of the proton~\cite{moments} was also used to evaluate $|G_E|$/$|G_M|$  and $|G_M|$. A comparison of the results obtained by the two methods is shown in Table~\ref{tab_ppbar1}. While the measurements by the different experiments concerning the EFF show very good agreement, this is not the case of $|G_E|$/$|G_M|$ , where the measurements by BaBar~\cite{BABARpp} and PS170~\cite{PS170} disagree for low $q$.

\begin{table}[h!]
\caption{Results on $|G_E/G_M|$ and $|G_M|$ by fitting the proton angular distribution as well as by using the method of moments at different c.m. energies. Reprinted from~\cite{xiaorong}.}
\label{tab_ppbar1}
\tabcolsep7pt\begin{tabular}{lccc}
\hline
$\sqrt{s}$~(GeV)  &$|G_E/G_M|$   & $|G_M|$~($\times10^{-2}$)  \\
\hline
        &{Fit on $\cos\theta_{p}$}  \\
2.2324  &$0.87\pm0.24\pm0.05$  &$18.42\pm5.09\pm0.98$  \\
2.4000  &$0.91\pm0.38\pm0.12$  &$11.30\pm4.73\pm1.53$  \\
(3.0500-3.0800) &$0.95\pm0.45\pm0.21$ &$3.61\pm1.71\pm0.82$ \\
        &{Method of moments}  \\
2.2324  &$0.83\pm0.24$    &$18.60\pm5.38$  \\
2.4000  &$0.85\pm0.37$    &$11.52\pm5.01$  \\
(3.0500-3.0800) &$0.88\pm0.46$     &$3.34\pm1.72$ \\
\hline
\end{tabular}
\end{table}

At present, the precision in the measurement of $|G_E|$/$|G_M|$ of the proton is dominated by statistics. Using the data collected by BESIII in 2015 (650  $\mathrm{pb^{-1}}$), statistical accuracies below $10\%$ are expected to be achieved. Considering some of the statistics collected at different charmonium and XYZ-states above 3.77 GeV  (7.4$~\mathrm{fb}^{-1}$), BESIII has higher ISR visible luminosity than BaBar (500$~\mathrm{fb^{-1}}$) for the study of the ISR channel $e^+e^{-} \rightarrow p \bar{p} \gamma$. This is due to the fact that BESIII can also perform untagged  analysis of these events (ISR photon emitted at very low or very large polar angles) for  $q >$  2.0 GeV. Comparable precision to BaBar is therefore expected in the measurement of the hadronic cross section and EFF from $e^+e^{-} \rightarrow p \bar{p} \gamma$ in BESIII. 

Experimental results on the annihilation cross section into $n\bar{n}$ are very scarce~\cite{dm1,dm2,fenice,snd}. Like in the case of the proton, two different measurements are possible in BESIII: $e^+e^{-} \rightarrow n \bar{n}$ and $e^+e^{-} \rightarrow n \bar{n} \gamma$. In both cases, while the $\bar{n}$-annihilation releases a large amount of energy (more than twice the neutron mass) in a broad hadronic shower in the EMC, neutrons do not produce such a broad shower and deposit much less energy. The sources of background in detecting $n$ and $\bar{n}$ are other neutral particles, like photons, neutral kaons and mostly beam background and neutral cosmic rays. In BESIII, the depth of the EMC amounts to 15$X_0$ or $0.7\lambda_I$, which means that the probability that a nucleon interacts with the EMC material is of about $50\%$. Since the energy resolution in the EMC is $2.5\%$ ($5\%$) for a 1 GeV photon in the barrel (endcaps), it is possible to at least reconstruct the $\bar{n}$-annihilation. The selection strategy could be based on the identification of the $\bar{n}$ and the ISR photon through shower shape analysis in the EMC and then detection of $n$ in the EMC with the help of the kinematics of the process. The detection efficiency of $\bar{n}$ vs photons can be extremely improved with the use of Multivariate Analysis Tools (Neural Networks). Also the use of other sub-detectors like the MUC and the TOF can be exploited such that the detection of the $n$ is not needed to identify $e^+e^{-} \rightarrow n \bar{n}$ and $e^+e^{-} \rightarrow n \bar{n} \gamma$ events. 
The probability of interaction of a nucleon inside the MUC is $96\%$. The TOF counters could be fired by the $\bar{n}$-annihilation, making the detection of $n$ unnecessary. Using the available scan data from 2015 and the data collected at different resonances above 3.77 GeV, BESIII can provide a measurement of the cross section of the Born annihilation process in a wide continuous $q$ range from $n\bar{n}$-threshold up to 3.08 GeV. Furthermore, not only the EFF can be extracted but also the ratio $|G_E|$/$|G_M|$ of the neutron could be extracted for the first time using BESIII 2015 scan data.

\begin{figure}[t]
\begin{minipage}{16.5pc} \hspace{1cm}
\includegraphics[width=16.5pc]{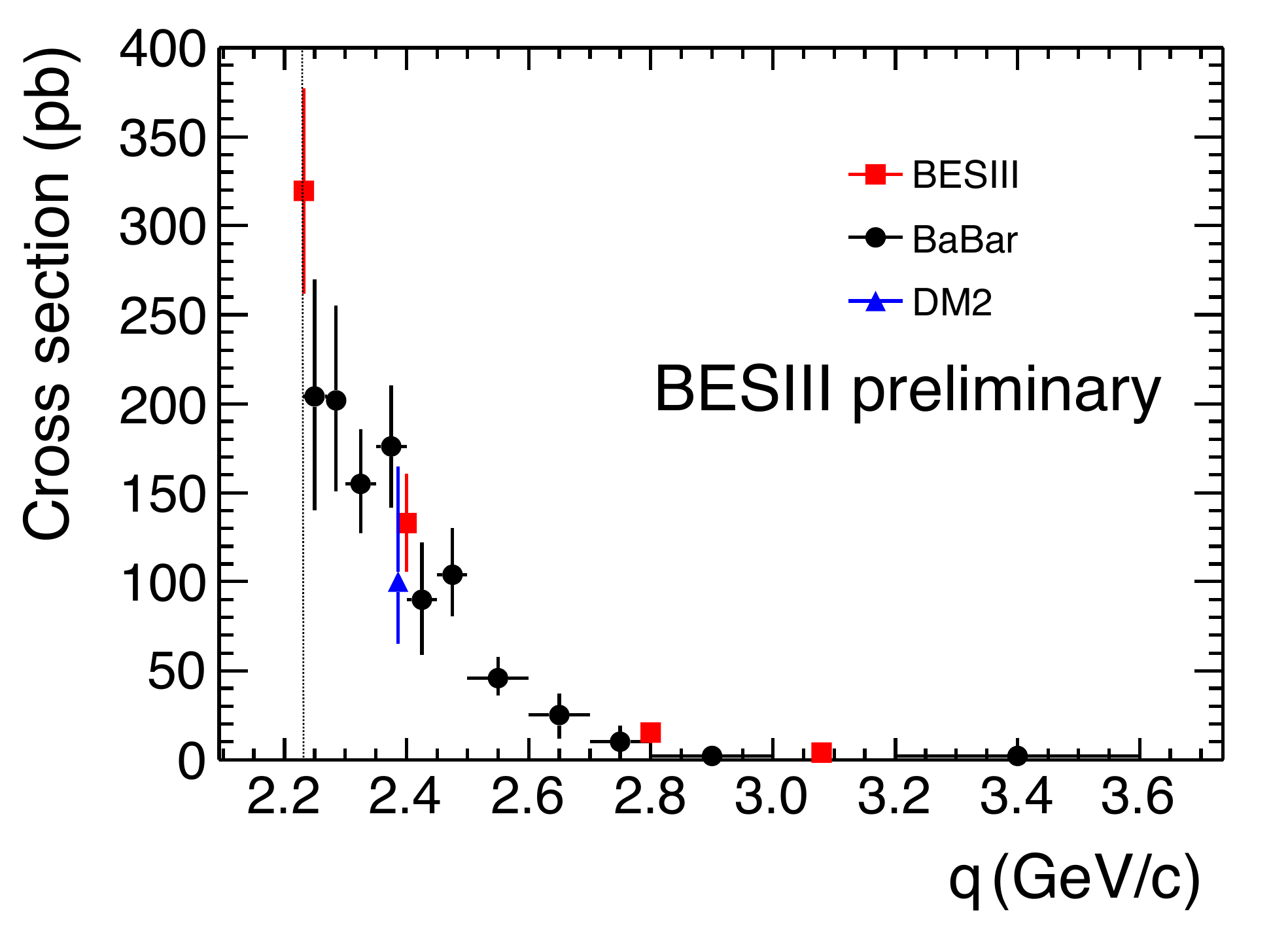}
\caption{dagv<bvy} \label{lambda}
\end{minipage}\hspace{2pc}%
\begin{minipage}{16pc}
\includegraphics[width=16.5pc]{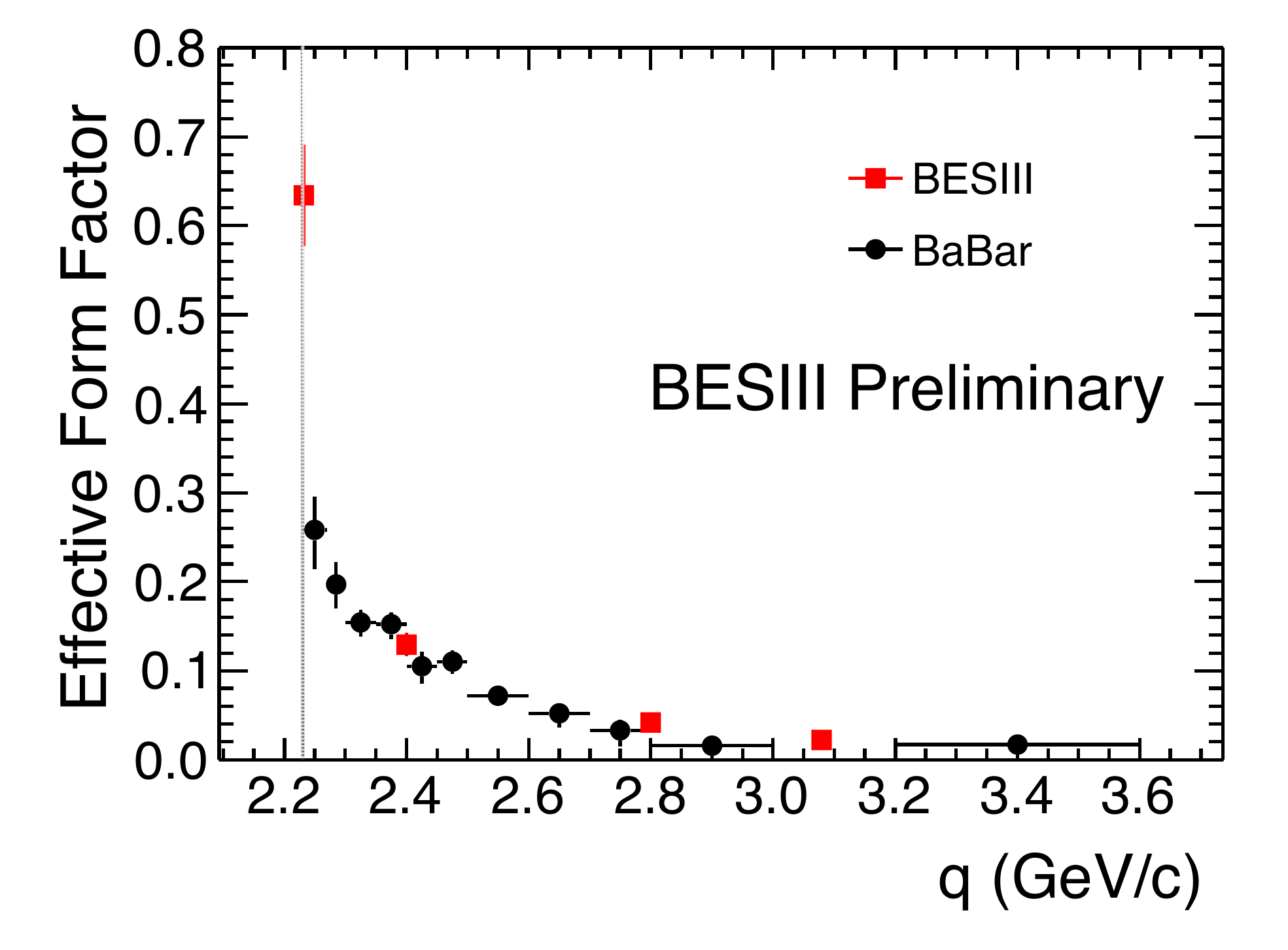}
\caption{Measurements of $e^+e^{-} \rightarrow \Lambda \bar{\Lambda}$ cross section (left) and $\Lambda$ effective form factor (right).} \label{lambda}
\end{minipage}
\end{figure}

\section{HYPERON FORM FACTORS}
BESIII has preliminary results on the measurement of the channel $e^+e^{-} \rightarrow \Lambda \bar{\Lambda}$. The analysis is based on $40.5~\mathrm{pb}^{-1}$ collected in 4 different scan points during 2011 and 2012. The lowest energy point is at 2.2324 GeV, only 1 MeV above the $\Lambda\bar{\Lambda}$-threshold. This makes it possible to measure the cross section almost at threshold. To use as much statistics as possible, both events where $\Lambda$ and $\bar{\Lambda}$ decay to the charged mode ($\mathrm{BR}(\Lambda \rightarrow p\pi^-) = 64\%$) and events where the $\bar{\Lambda}$ decays to the neutral mode  ($\mathrm{BR}(\bar{\Lambda} \rightarrow \bar{n} \pi^0) = 36\%$) have been selected. In the first case, the identification relied on finding two mono-energetic charged pions and a possible $\bar{p}$-annihilation.
In the second case, the $\bar{n}$-annihilation was identified through the use of Multivariate Analysis of EMC variables and additonally a mono-energetic $\pi^0$ was reconstructed to fully identify the channel. 
For the other 3 energy points, only the charged decay modes of $\Lambda$ and $\bar{\Lambda}$ were reconstructed by identifying the charged tracks and using the event kinematics. The preliminary results on the measurement of the Born cross section are shown in Figure~\ref{lambda} (left) together with previous measurements~\cite{LAMBDAS,LAMBDAS2}. The cross section at threshold is found to be $318 \pm 47 \pm 37$ pb, clearly different from zero. Given that the Coulomb factor in Eq.~\ref{diff} is equal to 1 for neutral baryon pairs, the cross section is expected to go to zero at threshold. This result confirms BaBar$^\prime$s measurement close to threshold~\cite{LAMBDAS2} with the difference that BESIII measurement is not integrated within a finite bin, like in the case of BaBar$^\prime$s  measurement through ISR. The BESIII measurement improves at least by 10$\%$ previous results at low $q$ and even more above 2.4 GeV. From the measurement of the cross section, the extraction of the EFF was also possible (Fig.~\ref{lambda} (right)). The cross section and EFF results are summarized in Table~\ref{tab_llbar}.

\begin{table}[h!]
\caption{Results on $e^+e^-\rightarrow\Lambda\bar{\Lambda}$ cross section and $\Lambda$ effective form factor measurements.}
\label{tab_llbar}
\tabcolsep7pt\begin{tabular}{lccc}
\hline
$\sqrt{s}$~(GeV)  &$\sigma_{Born}$~($pb$)   &$|G|$~($\times10^{-2}$) \\
\hline
2.2324    \\
$\Lambda \to p\pi^-,\bar{\Lambda} \to \bar{p}\pi^+$   &$325\pm53\pm46$ \\
$\bar{\Lambda}\to\bar{n} \pi^0$  &$(3.0\pm1.0\pm0.4)\times10^{2}$ \\
combined                  &$318 \pm 47 \pm 37$      &$63.2 \pm4.7 \pm 3.7$ \\\hline
2.4000    &$133\pm20\pm19$    &$12.9\pm 1.0\pm0.92$ \\
2.8000    &$15.3\pm5.4\pm2.0$ &$4.2\pm0.7\pm0.3$ \\
3.0800    &$3.9\pm1.1\pm0.5$  &$2.21\pm0.31\pm0.14$ \\
\hline
\end{tabular}
\end{table}

Using BESIII scan data from 2015, a full determination of the lambda FFs is possible. The imaginary part of the FFs leads to polarization observables. Since the $\Lambda$ decays through parity violation to proton and pion, the relative phase between $G_E$ and $G_M$, $\phi$, can be extracted from 
\begin{equation}
\label{polarization}
\frac{dN}{d\mathrm{cos}\theta_p} \propto 1 + \alpha_\Lambda P_n \mathrm{cos}\theta_p\hspace{0.75cm} \mathrm{and} \hspace{0.75cm}
P_n = - \frac{\mathrm{sin}2\theta_\Lambda \mathrm{sin}2\phi/\tau}{R\mathrm{sin}^2\theta_\Lambda/\tau + (1 + \mathrm{cos}^2\theta_\Lambda)/R} = \frac{3}{\alpha_\Lambda} <\mathrm{cos}\theta_p>, 
\end{equation}
where $N$ is the number of $\Lambda \rightarrow p \pi^-$ decays, $\theta_p$ is the angle between the p and the polarization axis of the $\Lambda$ in the $\Lambda\bar{\Lambda}$ c.m., $\alpha_\Lambda$ is the $\Lambda$ asymmetry, $P_n$ is the $\Lambda$ polarization, $\theta_\Lambda$ is the polar angle of the $\Lambda$ in the c.m., and $R$ is the ratio $|G_E|$/$|G_M|$ of the $\Lambda$. The same is also valid for the $\bar{\Lambda} \rightarrow \bar{p} \pi^+$ , therefore both decays can be added and used in the polarization measurement. The expected statistical accuracies from the analysis of 
$e^+e^{-} \rightarrow \Lambda \bar{\Lambda}$ from BESIII scan data in 2015 range between 6 and 17$\%$ and  for $|G_E|$/$|G_M|$ between 14 and 29$\%$. Similar measurements might also be possible in other hyperon channels like $e^+e^- \rightarrow \Lambda \bar{\Sigma}^0, \Sigma^0\bar{\Sigma}^0, \Sigma^+\bar{\Sigma}^-, \Xi^0\bar{\Xi}^0, \Sigma^-\bar{\Sigma}^+, \Omega^-\bar{\Omega}^+,\Lambda^+_c \bar{\Lambda}^-_c $.

\section{PION FORM FACTOR}
\begin{figure}[t]
\begin{minipage}{16pc} \hspace{1cm}
\includegraphics[width=16pc]{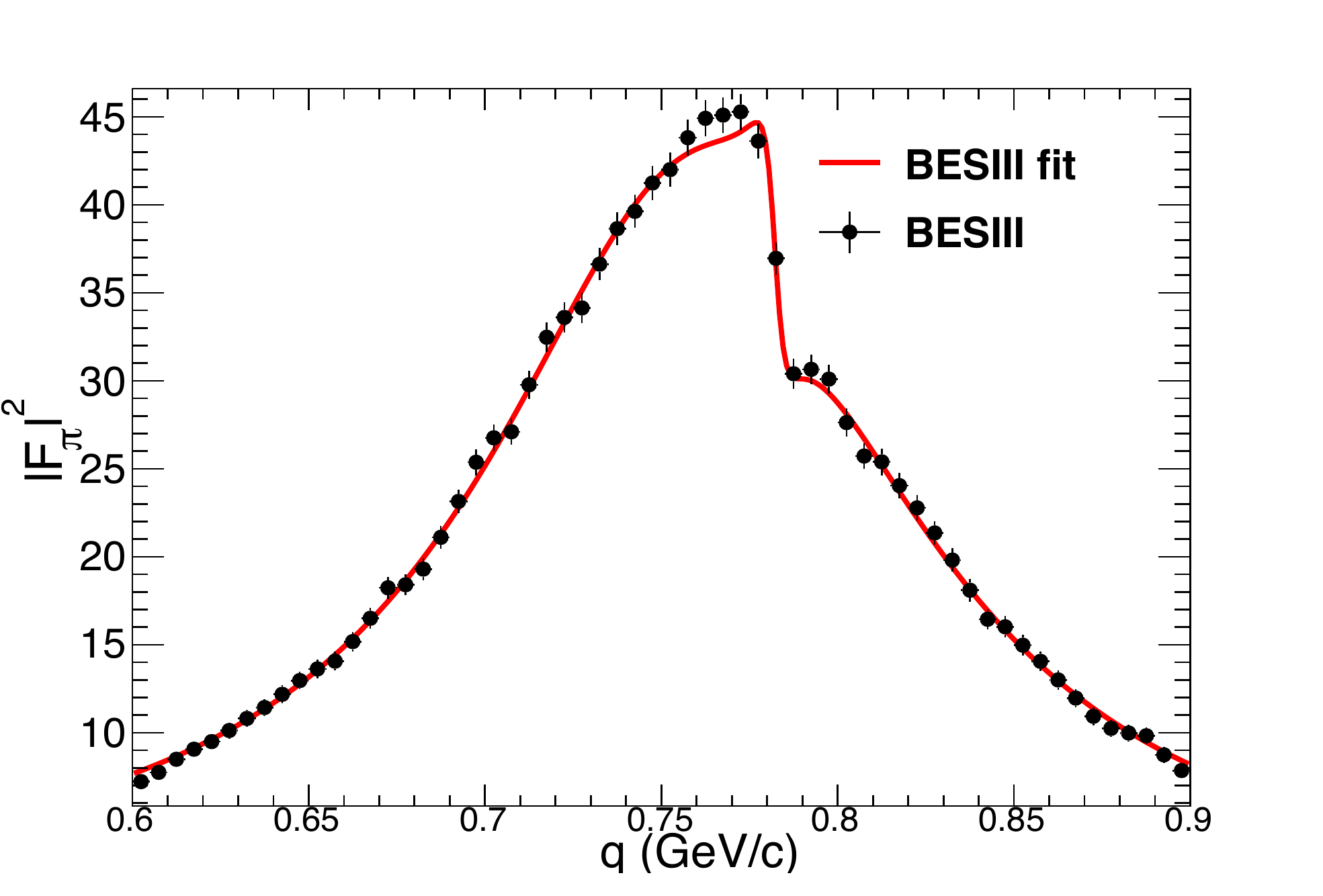}
\caption{dagv<bvy} \label{pion}
\end{minipage}\hspace{2pc}%
\begin{minipage}{17pc}
\includegraphics[width=17pc,height=11.pc]{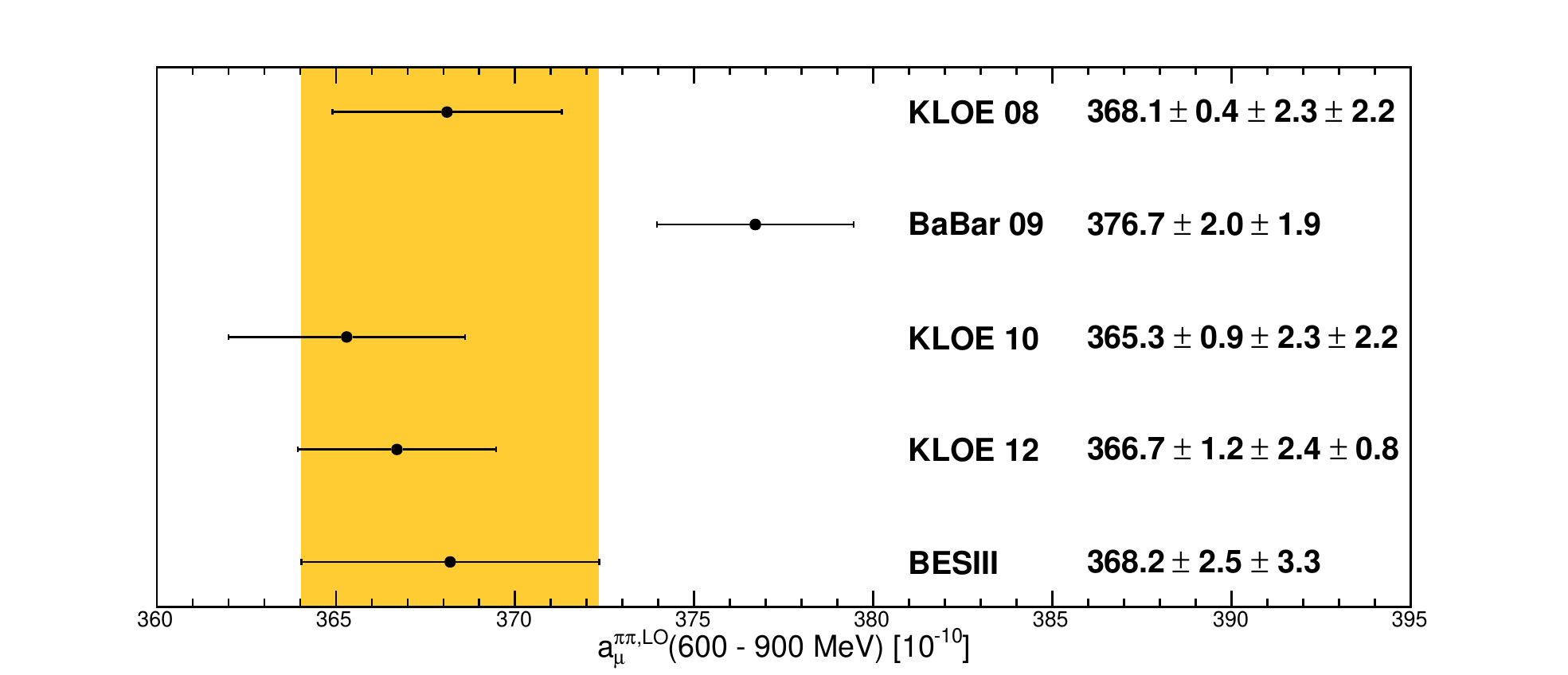}
\caption{Measured squared pion form factor and Gounaris-Sakurai fit parametrisation (left). Leading order hadronic vacuum polarization 2$\pi$ contributions to $(g-2)_{\mu}$ (right). Reprinted from~\cite{benedikt}.}
\end{minipage}
\end{figure}

Presently, the experimental and Standard Model (SM) values of the anomalous magnetic moment of the muon $a_\mu = (g -2)_\mu/2$ differ by more than 3 standard deviations~\cite{deviation,deviation2}. 
The SM calculation does not only receive contributions from QED, but also from weak and strong interactions and probably also from Beyond SM interactions.  In order to improve the SM predictions of $a_\mu$, the most important uncertainty is the contribution of the strong interaction. The hadronic contribution is split in two parts: the Hadronic Vacuum Polarization (HVP) and the Light-by-Light Scattering. The measurements of the hadronic cross section in $e^+e^-$-annihilation at low energies can be used to compute HVP by means of dispersion integrals. In particular, $e^+e^- \rightarrow \pi^+\pi^-$ contributes to more than 70$\%$ to the total HVP contribution. 
The ISR method was used by BESIII at 3.773 GeV (2.93 $\mathrm{fb}^{-1}$) to measure the channel $e^+e^- \rightarrow \pi^+\pi^-\gamma$ and access the $q$ region between 600 and 900 MeV, where the $\rho$ resonance is located~\cite{benedikt}. This energy range contributes more than 70$\%$ to the two-pion contribution to $a_\mu^{\pi\pi}$ and to about $50\%$ of the total hadronic vacuum polarization correction of $a_\mu$. The analysis was performed by tagging the ISR photon and identifying the pions in the final state. The selection criteria included the use of Multivariate Analyis to precisely separate pions from muons. For the extraction of the cross section Eq.~\ref{xsmeasured} was used, with $\epsilon$ the signal efficiency (including ISR), $\mathcal{L}$ the ISR luminosity,  and additionally a vacuum polarization factor.
Phokhara NLO generator~\cite{phok} was used to measure $\epsilon$ and the vacuum polarization factor. The Gounaris-Sakurai parametrization~\cite{sak} for the $\rho - \omega$ interference was applied to fit the $\pi^+\pi^-$ cross section and obtain the pion form factor (Fig.~\ref{pion} (left)). The obtained fit parameters agree well with existing results from PDG~\cite{pdg14} (Table~\ref{tabpion}).

\begin{table}[h!]
\caption{}
\label{tabpion}
\tabcolsep7pt\begin{tabular}{lccc}
\hline
Parameter & BESIII Value &PDG2015 \\
\hline
$m_\rho (\mathrm{MeV}/c^2)$    &$776.0 \pm 0.4$ & $ 775.26 \pm 0.25$ \\
$\Gamma_\rho$ (MeV) & $151.7 \pm 0.7$ & $147.8 \pm 0.9$ \\
$m_\omega (\mathrm{MeV}/c^2)$ & $782.2 \pm 0.6$ & $782.65 \pm 0.12$ \\
$\Gamma_\omega$ (MeV) & fixed to PDG & $8.49 \pm 0.08$ \\
$|c_\rho| (10^{-3})$  & $1.7 \pm 0.2$ & - \\
$|\phi_\omega|$ (rad) & $0.04 \pm 0.13$ & - \\
\hline
\end{tabular}
\end{table}
\noindent The bare cross section is needed as input for the calculation of $a_\mu^{\pi\pi}$. It is obtained from the previous cross section corrected for final state radiation (FSR). The FSR correction factor is determined with Phokhara NLO and theory~\cite{FSR}. The bare cross section and the pion form factor were measured with an accuracy of 0.9$\%$. The contribution of the cross section measurement to the hadronic contribution to $a_\mu$ was also calculated for the energy range between 600 and 900 MeV. As summarized in Fig.\ref{pion} (right), the BESIII result, $a_\mu^{\pi\pi, \mathrm{LO}}(600 - 900 \mathrm{MeV}) = (370.0 \pm 2.5_{\mathrm{stat}} \pm 3.3_{\mathrm{sys}})$, was found to be in between the corresponding values of previous measurements~\cite{Kloe1,BABARpi}. 

\section{SUMMARY}
BESIII is an excellent laboratory for the measurement of hadron FFs, since both ISR and scan measurements can be performed. Results on the measurement of proton FFs from scan data from 2011 and 2012 and pion FFs from ISR measurements at 3.773 GeV have been recently published. There are also preliminary results on lambda EFF based on scan data from 2011 and 2012. In 2015, BESIII performed a high luminosity scan ($650~\mathrm{pb^{-1}}$) in the region between 2.0 to 3.08 GeV with the aim to measure baryon FFs and the inclusive hadronic cross section, $R_{\mathrm{incl}}$, with unprecedented accuracy. Currently, BESII has more than $7.4 \mathrm{fb}^{-1}$ data above 3.77 GeV which could also be used for ISR measurements. The ISR visible luminosity of this data is comparable to BaBar$^\prime$s ISR visible luminosity.
\section{ACKNOWLEDGMENTS}
This work was supported  by German Research Foundation DFG under Collaborative Research Center CRC-1044.



\end{document}